# Human Resource Development and the Internet of Things

Robert M. Yawson*

School of Business,

Quinnipiac University, 275 Mount Carmel Ave, Hamden, CT 06518

ORCID iD: https://orcid.org/0000-0001-6215-434

robert.yawson@qu,.edu

Daniel Woldeab

College of Individualized Studies

Metropolitan State University, St. Paul, MN

Daniel.Woldeab@metrostate.edu

Emmanuel Osafo

Youth and Families Program Unit, Washington State University Extension

Washington State University, WA

ORCID iD: https://orcid.org/0000-0002-5385-3635

emmanuel.osafo1@wsu.edu

*Corresponding Author

**Abstract**

The Internet of Things (IoT) is affecting national innovation ecosystems and the approach of organizations to innovation and how they create and capture value in everyday business activities. The Internet of Things (IoT), is disruptive, and it will change the manner in which human resources are developed and managed, calling for a new and adaptive human resource development approach. The Classical Internet communication form is human-human. The prospect of IoT is that every object will have a unique way of identification and can be addressed so that every object can be connected. The communication forms will expand from human-human to human-human, human-thing, and thing-thing. This will bring a new challenge to how Human Resource Development (HRD) is practiced. This paper provides an overview of the Internet of Things and conceptualizes the role of HRD in the age of the Internet of Things.

## Human Resource Development and the Internet of Things

Since the launch of the world-wide web in the early 1990s, the Internet has impacted the way we live and work with the 'speed of light.' Society is facing yet another wave of Internet technologies that will have a big impact on the way we live and work. This phenomenon known popularly as the Internet of Things (IoT) presents a situation where data generation is the order of the day – every human interaction whether with living or non-living things generate some form of data making the workplace a data-driven environment. The changing phase of technology presents a challenge to predictions of human interaction and how work is conducted. The IoT has the potential to make a fundamental shift in the way we interact with our surroundings. "It is suggested that we can see the Internet as enabling the human social environment, as well as an ever-increasing array of Internet-enabled devices, to function as literal body parts" (Smart, 2017, p. 360).

Human Resource Development (HRD) is in a distinctive position to prepare the workforce for this new way of working and to utilize the big data generated by the Internet of Things (IoT). The IoT has the potential to fundamentally shift the way we interact with our surroundings (Manyika et al., 2015). The capability to monitor and manage objects in the physical world electronically makes it promising to bring data-driven decision-making to new realms of human resource development—to augment the performance of systems and processes, save time for people and businesses, and improve quality of life (Manyika et al., 2015).

Human Resource Development (HRD) involves developing people with a focus on improving knowledge, skills, and abilities (KSAs) to guide organizations, create a long-term vision, develop strategy, staff the organization, communicate, motivate people toward the vision, and to support improved productivity. HRD is targeted across levels of abstraction of individuals, teams, organizations, communities, and fields of policy and practice (Yawson, 2017). All these levels and focus of HRD are being impacted by the emergence of the IoT. As Bennett, (2014) aptly described:

> The field of HRD is at a historic point in which we can demonstrate value and relevance to the modern, technology-enabled organization. Many in the field of HRD have sought a balance between the needs of the individual and the needs of the collective for learning and performance. Both management and HRD needs are often embedded in the same virtual systems, but HRD has been late to incorporate technology strategically in practice and in academic preparation programs (p.275).

It is no longer business as usual for HRD professionals. As a result of the emergence of IoT, the world is experiencing significant, largely economic and sociotechnical, induced changes. These changes are more than jargon, cliché, and hyperbole, and they are effecting major transformations (Yawson & Greiman, 2014). These transformations will impact on how human resources are developed and we need to be able to forecast its effects (Yawson & Greiman, 2017). To produce such forecasts, HRD needs to become more predictive and adaptable - to develop the ability to understand how human capital systems, organizations, and the national innovation ecosystems will behave in the future that IoT brings. The Classical Internet has radically altered the way we access information, profoundly transformed the way we think, act and remember (Smart, Heersmink, & Clowes, 2017). With the IoT, every aspect of our cognitive and epistemic endeavors, either individual or collective, will be undertaken with some involvement of the Internet (Smart et al., 2017). Relative to this influence, it makes sense, to see the IoT as fully becoming an important part of the "cognitively-potent extra-organismic environment in which our biological brains are now situated." (Smart et al., 2017, p. 255). The

IoT can, therefore, be seen as a form of cognitive ecology that shapes our thinking and other socially transmitted ideas. Given the emergence, momentum, and the prospects of IoT, the objective of this research is to discuss the impact the advances in IoT will have on HRD research and practice and the role HRD should play in addressing the impact of IoT on the human resource in organizations.

**Research Question**

There is the need for optimal balance in modern core skills, like agility, collaboration,

cognitive flexibility, creativity and organizational development. It all comes down to educating and preparing the human resource across levels of abstraction of individuals, teams,

organizations, communities, and fields of policy and practice to absorb the big data that comes

from IoT. *Given this need as a result of the momentum and emergence of IoT, what role is there*

*for HRD as a field of study and practice to ensure success and relevance in this new era*?

**Research methodology**

To achieve the objective of this study, an integrated review of the literature was

performed. The reviewed literature included journal articles, conference papers, edited

volumes, and reports from several respected think tanks. Given that the IoT is still emerging, a

wide range of sources for a comprehensive review of the topic including the gray literature was

conducted. Given the nascent and fluidity of the emergence of IoT as a phenomenon, reviewing

only peer-reviewed scholarly articles that make a specific theoretical contribution to the IoT

would have yielded a very limited review. Despite the significant work done by members of the

Virtual HRD Special Interest Group of the Academy of Human Resource Development, most of the work is nascent and restricted to works published in *Advances in Human Resource Development* (Bennett, 2010; McWhorter, 2010; Nafukho, Graham, & Muyia, 2010).

Relevant literature was identified by querying scholarly databases for the terms: Internet of Things, IoT, Web of Things, Internet of Everything, Internet of Objects, Embedded Intelligence, Connected Devices and Technology Omnipotent, Cyber-Physical Systems, Pervasive Computing, Ubiquitous Computing, Machine-to-Machine; Human-Computer Interaction, and Ambient Intelligence. Each of these terms were searched in combination with term Human Resource. Returned results were downloaded and uploaded to Mendeley Citation Software and further screened using the following terms: Human Capacity and Human Resource. The scholarly databases queried included: Web of Science, ABI/INFORM Global, Academic Search Premier, ACM Digital Library, Applied Science & Technology Full Text (EBSCO), IEEE Xplore, ScienceDirect, and Google Scholar. The resulting 47 pertinent articles were reviewed for the study.

## The of Internet of Things

The concept of combining computers, sensors, and networks to monitor and control devices is not new. It has been around for several decades. However, the recent confluence of key technologies like microelectronics, nanotechnology, biotechnology, cognitive sciences, synthetic biology, Information Communication Technologies (ICT), and market trends is ushering in a new reality for the IoT.  IoT promises to usher in a revolutionary, fully interconnected "smart" world, with relationships between objects and their environment and objects and people becoming more tightly intertwined (Rose, Eldridge, & Chapin, 2015). The vista of the IoT as a pervasive array of devices bound to the Internet will fundamentally change how people think about what it means to be "online" (Rose et al., 2015). Technically, the IoT is not the result of a single novel technology; instead, several complementary technical developments and innovations provide systemic capabilities that help to bridge the gap between the virtual and physical world (Mattern & Floerkemeier, 2010).

**The Episteme and Taxonomy of the Internet of Things**

Kevin Ashton is credited with coining the term "Internet-of-Things" in a presentation in

1999 regarding supply chain management (Ashton, 2009; Gubbi, Buyya, Marusic, &

Palaniswami, 2013). Since then, Internet of Things (IoT) has emerged as a new paradigm aiming

at providing solutions for integration, communication, data consumption and analysis of smart

devices (Khodadadi, Dastjerdi, & Buyya, 2017). While the term "Internet of Things" is relatively

new, the concept of joining computers and networks to monitor and control devices has been

around for decades (Rose et al., 2015). The story of the IoT can be traced back to the 19[th]

century when the electromagnetic telegraph was created by Baron Schilling in Russia (Borisovai,

2009). Figure 1 is an illustrative flow diagram depicting the history and evolution of the IoT. The

evolution saw major landmarks including innovations, significant events, thoughts and

predictions from pioneers and thinkers. In an interview with *Colliers Magazine* in 1926, Nikola

Tesla stated:

> When wireless is perfectly applied, the whole earth will be converted into a huge
> brain, which in fact it is, all things being particles of a real and rhythmic whole, and
> the instruments through which we shall be able to do this will be amazingly simple
> compared with our present telephone. A man will be able to carry one in his vest
> pocket. (Kennedy, 1926, Webpage)

Figure1: The History and Evolution of the Internet of Things

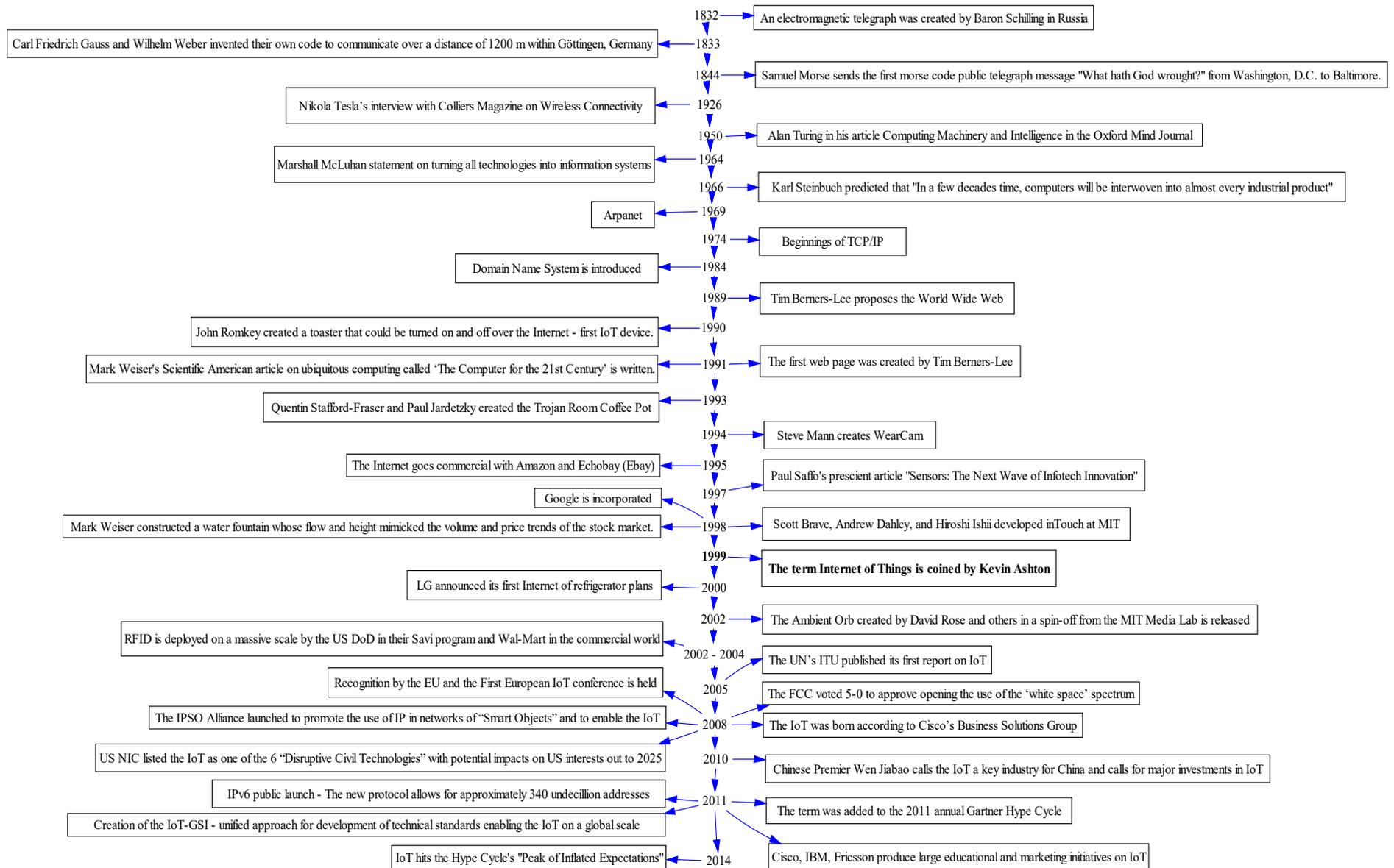

Source:   © Authors

In his 1964 book, *Understanding Media,* Marshall McLuhan stated that "....by means of electric media, we set up a dynamic by which all previous technologies - including cities - will be translated into information systems" (McLuhan, 1964). In 1966, Karl Steinbuch, a German computer science pioneer also predicted that "In a few decades time, computers will be interwoven into almost every industrial product" (Mattern & Floerkemeier, 2010, p. 242)

The World Wide Web (Web 1.0) - a network of linked HTML documents that resided on top of the Internet architecture - characterized the early days of the Classical Internet – the Internet as we know it today. This network of static HTML pages progressively evolved into Web 2.0, a term describing the use of World Wide Web technology and web design that enabled creativity, secure information sharing, collaboration and functionality of the web. With Web 2.0, two-way communication became ubiquitous and allowed user participation, collaboration, and interaction (Whitmore, Agarwal, & Da Xu, 2015). Web 2.0 technologies include social networking services, electronic messaging services, blogs, and wikis—technologies that have become indispensable to modern social interaction as well as for global business.

While Web 2.0 currently dominates the Internet, there has been the emergence of Semantic Web or Web 3.0. A technology that makes markup web content understandable by machines, allowing machines and search engines to behave more intelligently (Whitmore et al., 2015). Marking up web content in standardized formats would allow machines to process and share data on their own, without the need for human mediation (Whitmore et al., 2015). Alongside developments in the Internet technologies, technologies in Sensor Networks, Near

Field Communication using RFID tags, synthetic biology, biotechnology, cognitive sciences, and nanotechnology have also been evolving. Convergence of Web 2.0, Web 3.0, and these technologies, has led to a paradigm being referred to as the Internet of Things (IoT).

IoT is maturing and continues to be the latest, a most hyped concept in the IT world. It was added to the 2011 annual *Gartner Hype Cycle* that tracks technology life-cycles from "technology trigger" to "plateau of productivity," and it hit the Hype Cycle's "Peak of Inflated Expectations" in 2014. As of August 2017, the term IoT was still at the "Peak of Inflated Expectations". Gartner's Information Technology Hype Cycle (Gubbi et al., 2013) is popularly known for representing emergence, adoption, maturity, and impact on applications of specific technologies (Ferguson, 2002). It was forecasted in 2012 that IoT would take between 5-10 years for market adoption and every indication now is evident it was predicted right.

Riggins and Wamba, (2015) grouped the level of IoT adoption through Big Data analytics usage to the following categories:

1) *Society level* where IoT mainly influences and improves government services by reducing cost and increasing government transparency and accountability,

2) *Industry level* in which manufacturing, emergency services, retailing, and education have been studied as examples,

3) *Organizational level* in which IoT can bring the same type of benefits as those mentioned in society level,

4) *Individual-level* where daily life improvements and individual efficiency and productivity growth are marked as IoT benefits.

The IoT has been referred to with different terminologies, but the objective of IoT is same in the broad sense (Madakam, Ramaswamy, & Tripathi, 2015). The taxonomical labels of IoT include Internet of Everything, Web of Things, Internet of Objects, Embedded Intelligence, Connected Devices and Technology Omnipotent, Omniscient and Omnipresent. In addition to these taxonomical labels, IoT has also been variously described as follows (Madakam et al., 2015):

- *Cyber-Physical Systems*: Integrations of computation and physical processes, in which bringing the real and virtual worlds together.

- *Pervasive Computing*: A computer environment in which virtually every object has processing power with wireless or wired connections to a global network

- *Ubiquitous Computing or Calm technology*: Where technology becomes virtually invisible in our lives

- *Machine-to-Machine Interaction*: Means no human intervention while devices are communicating end-to-end

- *Human-Computer Interaction*: Involves the study, planning, and design of interaction between people and computers

- *Ambient Intelligence*: It is a developing technology that will increasingly make our everyday environment sensitive and responsive.

There are varying definitions of IoT, and there is not a standard one agreed to by all. However, there is a common understanding of what it is and its prospects. "What all of the definitions have in common is the idea that the first version of the Internet was about data created by people, while the next version is about data created by things" (Madakam,

Ramaswamy, & Tripathi, 2015, p.165). The *thing* in IoT can be a person with a heart monitor implant, a farm animal with a biochip transponder, a crop with a nanochip for precision agriculture, an automobile that has built-in sensors to alert the driver when the tire pressure is low—or any other natural or man-made object that can be assigned an IP address, and provided with the ability to transfer data over a network (Shin, 2014).

Fundamentally, the IOT can be described as a global network which facilitates the communication between human-to-human, human-to-things, and things-to-things, which is anything in the world by providing a unique identity to every object (Aggarwal & Das, 2012). Madakam et al., (2015) define IoT as "An open and comprehensive network of intelligent objects that can auto-organize, share information, data, and resources, react and act in the face of situations and changes in the environment" (p. 165)

*Figure 2*. The Quantum of Internet of things

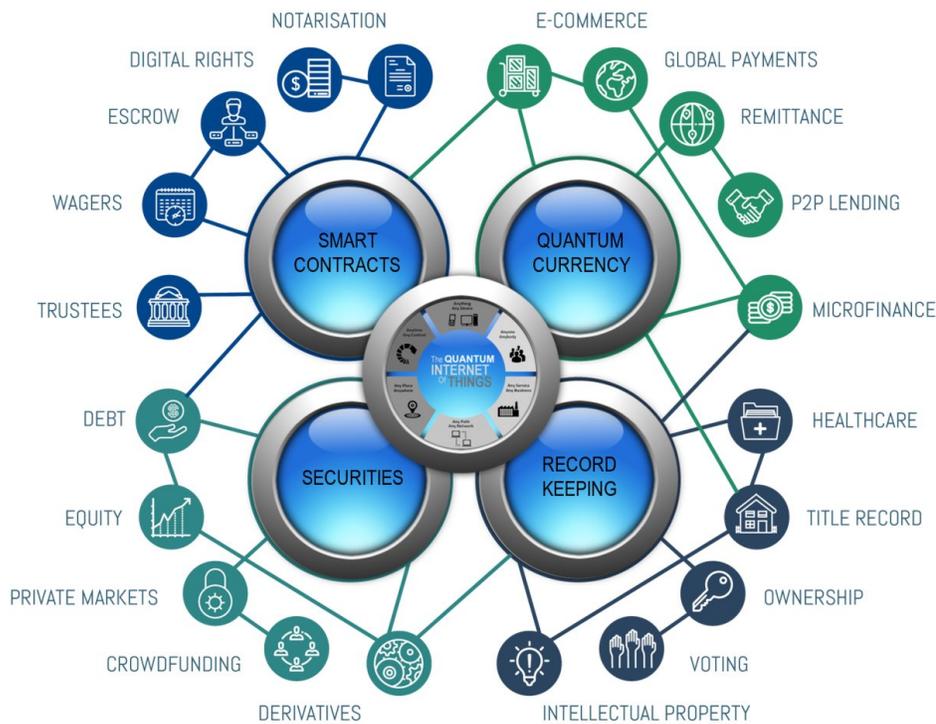

*Source: Mobile Analytics/Big Data by Joel Comm. Creative Commons License*

The Internet of Things is an emerging technological concept of sociotechnical and economic significance. Applications of IoT is a fast-growing segment of business communities worldwide, empowering industries globally by transforming their operation and giving them faster and more efficient ways of doing business. Consumer products, medical devices, pharmaceutical products, agrifood products, durable goods, cars and trucks, industrial and utility components, sensors, and other everyday objects are being linked with Internet connectivity and powerful data analytic capabilities that promise to transform the way we work, live, and play (Rose et al., 2015). In addition to interconnectivity and interoperability, the quantum of Internet of things is also very significant.  Joel Comm (2017) has predicted how quantum entanglement will impact how everyday business is conducted. Figure 3 is an illustration of the quantum IoT. The Quantum Internet of Things enables new ways of monitoring and managing all the "moving parts" that make up a business.

In their report "Unlocking the Potential of the Internet of Things'', the McKinsey Global Institute describes the broad range of potential applications regarding "settings" where IoT is expected to create value for industry and users (Manyika et al., 2015). "Some of the most promising uses are in health care, infrastructure, and public-sector services—helping society tackle some of its greatest challenges" (Manyika et al., 2013, p. 51). The Internet of Things is still in early stages of adoption, but it already has a wide variety of uses, and the portfolio of applications is expanding daily.

Indeed, in a world suffused with smart devices, it is not only our homes and workplaces that are changing but our way of life as figure 2 aptly depicts. However, smart objects are only the first step of an evolutionary process of IoT. There is a generational evolution from objects with a certain degree of smartness to objects with an actual social consciousness (Atzori, Iera, & Morabito, 2014). In analogy with the human evolution from *homo sapiens* to *homo agens* used in economic and sociological constructs, Atzori et al. (2014) used a similar evolutionary path from a *res sapiens* (smart object) to what they called *res agens* (an acting object), "which is able to translate the awareness of causal relationships — the basis of knowledge of change and evolution of its environment — into actions" (Atzori et al., 2014, p. 98).

*Figure 3.* Main features of the identified three categories of IoT objects

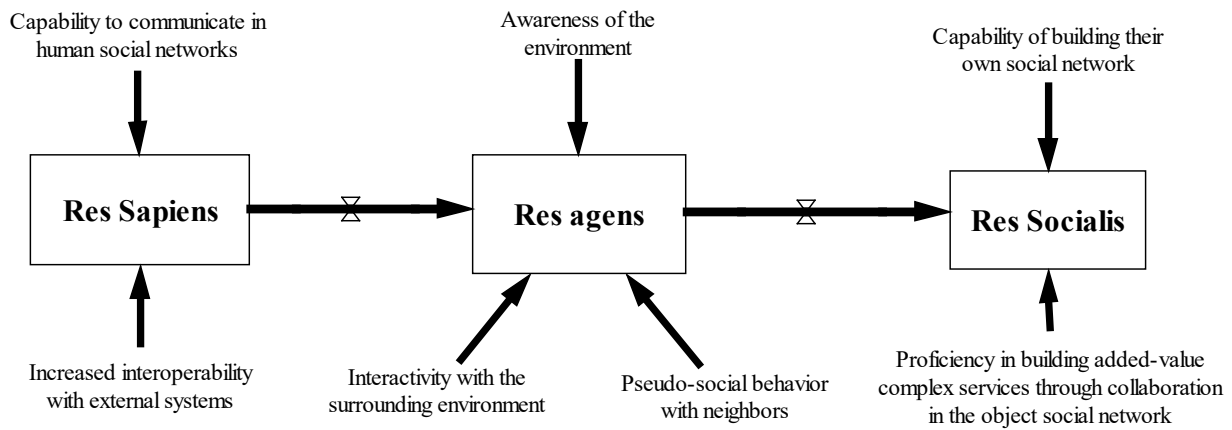

*Source: Adapted from Atzori et al. (2014)©Authors, 2018*

Atzori et al., (2014) went further on their analogical evolution ladder toward a new type of object that can be considered as *res socialis* (i.e., social object). The term they described as an object that is part of and acts in a social community of objects and devices (which, in this case, is a social IoT). The features of the three categories of IoT objects identified by Atzori et al. (2014) are illustrated in Figure 3.

The disruptive nature of IoT technologies and the fast-growing applications is evident in industries from manufacturing to pharmaceuticals and from health care, to our smart homes and workplaces (Atzori et al., 2014; Hoontrakul, 2018; Xu, He, & Li, 2014). Even livestock industries are using these technologies to keep track of their precious assets (Gao & Bai, 2014).

In the age of IoT, all of these widely diverse industries are gaining greater efficiency and exponential growth at rapid rates.

Undeniably, these changes in the workplace will have a great impact on education in general and higher education in particular. At the very least, such developments and evolutions in the workplace demand a more educated population. While DeMillo (2011) pointed out that technological advancements, in general, have the potential to be challenging to higher education institutions, Tianbo (2012) asserted that IoT is encouraging changes in higher education and has the potential to create more intelligent systems in these institutions. However, higher education institutions not only have to think about the transformative effects of IoT's emergence in our society, but are also obligated to rethink how best to educate the coming generation (Kortuem, Bandara, Smith, Richards, & Petre, 2013). Crucially, there is also "need for an education provision that can empower a new generation of a digital citizen who can understand both technologies that underpin the Internet of Things, as well as the societal impacts of widespread adaption of these technologies" (Kortuem et al., 2013, p. 53).

Anticipated impact of IoT on the Internet and economy are staggering, with some projecting as many as 100 billion connected IoT devices and a global economic impact of more than $11 trillion by 2025 (Rose et al., 2015). This is having very significant impact on organizations and human resources, and the role HRD will play may be dramatically altered.

**Discussions**

The disruptive potential of IoT could drive profound changes across many dimensions—in the lives of individuals, in business, and across the global economy. The IoT is such a sweeping concept that presents a challenge about how to imagine all the possible ways in which it will affect businesses, economies, and society (Manyika et al., 2013). However, it is undeniable that it will have a significant impact and all aspect our lives and the way human resources should be developed. Since 2010, when a special issue on Virtual HRD (VHRD) was published in the *Advances in Human Resource Development*, there has been an increasing stream, albeit, slowly the literature on technology-enabled HRD. However, there has not been any serious discussion or research as to the role HRD will play in the age of IoT and the impact of IoT on the practice of HRD as a profession in the mainstream HRD discussions.  It is, therefore, gratifying to learn that the keynote speaker for the 2018 Academy of Human Resource Development International Research Conference is a recognized expert and thought-leader in the adult development arena and will be discussing "Re-thinking learning and HRD for the Age of Artificial Intelligence."  As, authors we share the same view as the Keynote Speaker [Pat McLagan] that "learners' capabilities, confidence, and self-image in learning need a radical change to thrive in a world of proliferating resources, information overload, artificial intelligence, and many other paradigm-shifting forces."(AHRD, 2018, p.1)

Bennett, (2014) defines Virtual Human Resource Development (VHRD), as a "media-rich and culturally relevant webbed environment that strategically improves expertise, performance, innovation, and community-building through formal and informal learning" (p.265). While we agree with this definition, we contend that there should not be anything like VHRD. While this may sound controversial, our reason is that in the age of IoT the entirety of HRD including its connected yet disparate areas of scholarship like Strategic HRD, Critical HRD, International/Global HRD, etc., are all covered in the description of VHRD provided by Bennett, (2014). We can no longer see VHRD as a separate area of study under HRD but the new paradigm of HRD or the evolution of HRD. Bennett and Bierema, (2010) stated that "VHRD can be viewed as a living system because of the interactivity, learning, and development that occurs through its enabling technologies." We argue that the entirety of HRD in an IoT era should be viewed as such. HRD professionals need to identify potentially disruptive technologies, and carefully consider their potential before these technologies begin to exert their disruptive powers in the workplace and society. The impacts of IoT on HRD are considerable. The advances in IoT create a demand for new sets of skills, and as working adults assume these jobs, they need to be retrained and reskilled. Training and preparing the human capital needed to fill the high demand of high-tech jobs is going to be a considerable undertaking, which makes the implications for HRD enormous.

During the last few decades, millions of individuals globally have been raised from poverty into the middle class, which means they not only need but demand access to higher education (Kortuem et al., 2017). Hence, in the age of IoT, Adult Education and HRD are uniquely positioned to provide the education and training needed not only by the workforce of today but also that of tomorrow, which will face increasingly high-tech and shifting demands in the workplace.

We envisage the impact of IoT on HRD and the role HRD should play in maximizing the benefits and challenges of IoT under three dimensions of IoT applications to:

- *Inform* – gathering information through sensors to inform policy and HRD decision making, research and practice.

- *Automate* – design and develop activities by allocating a function to a system or by supervising the fulfillment of activity through an IoT device that can generate Big Data for effective predictions for the future.

- *Transform* – redesigning learning and development processes to moderate the distractions of the IoT.

**Implications for HRD**

Although Internet of Things provides HRD with tremendous opportunities for growth, it is not without challenges. In addition to its potential for enormous economic impact, the IoT will affect the performance of a range of organizations and individuals. As every aspect of our lives becomes ever more connected, thousands of discrete data points are created by just a handful of individuals on any given day. This provides an environment conducive to hackers and cyber criminals trying to gather sensitive employee and consumer information. When the emergence of new technology outpaces security developments, the likelihood that IoT can cause security and privacy breaches for HR practitioners is great.  After all, consumer data is one of the most precious assets of any organization, and assuring the security and privacy of this data in the age of IoT is imperative.

Schramm (2014)noted that IoT influence on HR is wide-reaching, from how data about workers are gathered and analyzed, to recruitment and employees' safety. Schramm (2014) also asserted that the data results gained through IoT today would help inform and influence HR in the future. Hence, HR practitioners should carefully balance the gains and opportunities that IoT presents, with the potential security privacy issues relating to employee and other data.

As technology shapes and amplifies our culture, and therefore our lives, it is vital for HR practitioners to understand and address the implications that IoT can have in the context of different cultures. The workforce of today is global. As such, HR/HRD practitioners should also understand the implications of different symbols and meanings, and how differently they can be perceived and interpreted by various cultures.

There are important implications for all stakeholders—consumers, IoT user companies, technology suppliers, policymakers, and employees. Catching up with the velocity at which the IoT affect work design and task performance can be challenging, but taking no action will result in virtual death. As much as the emergence of IoT has presented a challenge to relatively new and emerging fields like HRD it has as well created an opportunity for the ingenious to take a leadership role and announce their relevance in a new way. We perceive two main challenges that the IoT presents to HRD:

1. How to absorb the big data generated through the IoT – If the existing HRD systems struggle to absorb actionable workforce analytics data to understand how history informs the future, the enormity of the big data generated by IoT can subvert the field if immediate action is not taken. HRD can create an enabling environment for research in an IoT workplace by developing a cutting-edge database that enables continuous interaction between HRD scholars around the globe about their current research aside those presented at conferences. This strategy will help HRD scholars to identify others in the field whose research aligns with their thoughts and research ideas and serve as a catalyst for collaboration and exchange of ideas for enriched scholarly work that makes a greater impact on society. Furthermore, data from such a database can help students of HRD and emerging scholars to be informed about ongoing trends in the field and save them from the needless uncertainty about the relevance of their thesis in their academic/scholarly pursuit.

2. How HRD can use the IoT to connect workforce development to the people analytics principle –The use of technology and statistics to collect and analyze data to help management make informed decisions on talent acquisition and development is becoming more challenging as virtual training and assessment become increasingly popular. Thus, some people can fake their presence at such training and assessment sessions or cheat with technology. This calls for HRD professionals to build a strong network to help generate a comprehensive talent data from a variety of locations and issues to help improve our work. Professionals can share information constantly and make recommendations to other professionals through a well-develop database and HRD Collaboratories (Yawson, 2009). Linkedin and other organizations have taken the lead, but HRD as a field can improve upon their idea by having professionals in the different locations contribute to one another's work.

**Conclusion**

The emergence of IoT is undisputed and - as a phenomenon – irresistible (Maarten Botterman, 2015). IoT is disruptive, and it is driven by societal needs and economic opportunities; by demand pull and supply push. It is enabled by many different strands of technology innovation application development in domains as varied as synthetic biology, biotechnology, cognitive sciences, and nanotechnology. Instances are already seen in almost every area; the influence of IoT devices, services and architectures may rapidly become pervasive. These different forces certainly produce transitory conflicts of interests, gaps, and

distortions for which trade-offs need to be made (Maarten Botterman, 2015). Whether we as

HRD professionals recognize and respond to the IoT as a "thing in itself" will greatly influence the effectiveness of our ability to exploit and eventually resolve these tensions.

The IoT poses profound challenges to HRD as a field of study. Many stem from the

means in which it is likely to affect and even disrupt areas either of traditional HRD research

and practice (Performance Improvement, Training & Development, Leadership & Career

Development, Workplace Learning, etc.) or new frontiers of HRD like Knowledge Management,

Critical HRD, and International/Global HRD. For example in the area of CHRD, policy challenges

are arising from the IoT itself that will affect social justice, gender, and other related issues in

the domain of CHRD. In International/Global HRD, some of the issues would be similar to the

experience of other emergent technologies, especially those with the potential to transform

public services and the national innovation ecosystems. These include the need for suitable

forms of cross-cultural understanding (when things of different geographic and cultural

orientation are communicating with each other), access to skills and fair and efficient market

access. They also include organizational capital and human resource needs, "in particular for

business, entrepreneurial, technological and societal knowledge available to new and existing

enterprises moving into this area or building new businesses with the aid of IoT capabilities"

(Maarten Botterman, 2015, p. 26). Underpinning these is the need for a predictive and adaptable HRD research and practice capable of providing the right mix of certainty and flexibility.